\documentclass[aps,pre,twocolumn,float]{revtex4}
\usepackage{amsmath,bm,epsfig}

\let \nn  \nonumber

\let\*\cdot
\def\<{\left\langle} \def\>{\right\rangle} \def\({\left(} \def\){\right)}
 \let\~\widetilde \let\^\widehat 

\def\be{\begin{equation}}\def\ee{\end{equation}}
\def\bea{\begin{eqnarray}}\def\eea{\end{eqnarray}}
\def\bse{\begin{subequations}}\def\ese{\end{subequations}}
\newcommand{\BE}[1]{\begin{equation}\label{#1}}
\newcommand{\BEA}[1]{\begin{eqnarray}\label{#1}}
\newcommand{\BSE}[1]{\begin{subequations}\label{#1}}

\let \nn  \nonumber
\usepackage{pstricks}
\usepackage{pst-node}
\usepackage[ansinew]{inputenc}
\usepackage{amssymb,amsmath}

\def\BSE{\begin{subequations}}\def\ESE{\end{subequations}}
\let \= \equiv

\def\a{\alpha}
\def\b{\beta}

\def\o{\omega}

\def\be{\begin{equation}}       \def\ba{\begin{array}}

\def\ee{\end{equation}}         \def\ea{\end{array}}

\def\bea {\begin{eqnarray}}      \def\eea {\end{eqnarray}}

\def\bean{\begin{eqnarray*}}    \def\eean{\end{eqnarray*}}

\def\e{\varepsilon}           

\def\const {\mathop{\rm const}\nolimits}

\def\RA {\ \Rightarrow\ }

\def\<{\langle} \def\({\left(}  \def\>{\rangle} \def\){\right)}

\newtheorem{exi}{Example}

\begin{document}

\title{What can go wrong when applying  wave turbulence theory}
\author{Elena Tobisch\\
Institute for Analysis, Johannes Kepler University, Linz, Austria, \\ \emph{and} \\
Kavli Institute for Theoretical Physics, Santa Barbara, California, USA}

  \begin{abstract}
Many new models of wave turbulence -- frozen, mesoscopic, laminated, decaying, sand-pile, etc. -- have been developed in the last decade aiming to solve problems seemingly  not solvable in the framework of the existing wave turbulence theory (WTT). In this Letter we show that very often the reason of these discrepancies is that some necessary conditions of the WTT are not satisfied:  initial energy distribution is not according to the assumptions of the theory; nonlinearity is not small enough; duration of an experiment is not sufficient to observe kinetic time scale; etc. Two alternative models are briefly presented which can be used to interpret experimental data, both giving predictions at the dynamical time scale: a) a dynamical energy cascade, for systems with narrow initial excitation and weak and moderate nonlinearity, and b) an effective evolution equation, for systems with distributed initial state and small nonlinearity.
\end{abstract}
\maketitle

\section{Introduction}
The pioneering paper of Zakharov and Filonenko, \cite{zak2}, laid in 1968 the very foundations for the theory of weak (or wave) turbulence (WTT) in  systems with distributed initial state. Current developments of the WTT have been summarized in 1990, \cite{ZLF92}, where the general method for deducing the wave kinetic equation was presented, together with the method of finding its stationary solutions called kinetic energy spectra (K-spectra). K-spectra in a system with dispersion relation of the form $\o(k) \sim k^{\a}, \, \a>1$ satisfy a power law scaling, $k^{\nu}$. Here $k=|\mathbf{k}|$ is the modulus of the wave vector and exponent $\nu$ is a constant as soon as the dispersion relation $\o=\o(\mathbf{k})$ is known. Formation of the energy spectra takes place on the kinetic time scale described in detail hereinafter. Presently this theory is called \emph{kinetic} WTT, to be distinguished from \emph{discrete} WTT.

The discrete WTT has been developed for weakly nonlinear systems with localized initial state. In this case the total energy of the system is captured by a few independent clusters of resonantly interacting waves, without energy flow  between the clusters, on the dynamical time scale. The main properties of these clusters have first been described by Kartashova in the 1990s, \cite{PHD1,PHD2,PRL}. The concept of "discrete WTT" came to use in \cite{K06-1} and was explained in detail in \cite{K09b}. The current state of the art in discrete WTT is summarized in \cite{CUP}.

In the last decade many new models of WTT were developed, e.g. laminated \cite{K06-1}, frozen \cite{Pu99}, mesoscopic \cite{zak4}, decaying \cite{OOSRPZB}, sand-pile model \cite{Naz06}, etc. to solve problems seemingly  not solvable in the framework of the existing (kinetic and discrete) WTT.

In this Letter we present  studies of cases taken from literature showing  that very often the reason for these discrepancies is not that the WTT should  be improved. It is rather that some necessary conditions of the WTT are not fulfilled and the theory can not be applied at all. When possible,  we will also give references to some other theories or models which are more appropriate  for application in these specific cases.

Three main sources of misunderstanding are listed below:\\
$\bullet$ any prediction of  the WTT is comprised of two elements: a phenomenon (e.g. energy spectrum) and a time scale for the phenomenon to occur; very often the spectrum is taken neglecting the time scale;\\
$\bullet$ a dispersion function of decay type does not guarantee that we have a 3-wave system;\\
$\bullet$ in contrast to exact resonances, quasi-resonance clustering depends on both geometry and dynamics.

As a starting point we give some definitions. Regard a weakly nonlinear dispersive evolutionary PDE
\be \label{NEPDE}\mathcal{L}(\psi)=-\e \mathcal{N}(\psi)\ee
 such that $\mathcal{L}(\phi)=0,$
for any  $\phi=A_\mathbf{k} \exp {i[\mathbf{ k} \, \mathbf{x} - \o t]}$ with constant $A_\mathbf{k},$ wave vector $\mathbf{k}$  and dispersion function $\o=\o(\mathbf{k}), \, \o(\mathbf{k})^{''}\neq 0$.
Resonance conditions and dynamical systems (canonic variables, 3-wave resonances, 3WR) read
 \bea
\omega ({\bf k}_1) + \omega ({\bf k}_2)= \omega ({\bf k}_{3}), \ \
{\bf k}_1 + {\bf k}_2 ={\bf k}_{3}; \, \label{3R}\\
i\dot{B}_1=  Z B_2^*B_3,\,
i\dot{B}_2=  Z B_1^* B_3, \, i\dot{B}_3=  - Z B_1 B_2; \, \label{3D}
\eea
and the kinetic equation for this system (3WKE) takes the form
\bea
\frac{\bf d }{{\bf d } T}<{B}^2_3> =
 \int |Z|^2
\delta(\o_3-\o_1-\o_2)\delta({ k}_3-{ k}_1-{ k}_2) \nn \\
\cdot (B_1B_2-B_1^{*}B_3-B_2^{*}B_3)
{ d}{ k}_1 { d}{ k}_2; \, \label{3K}
\eea
where $B_i=B_i(T) \neq \const, \, T=t/\e $, are slowly changing amplitudes depending on  $T$.

\section{Time-scale}
\emph{The fundamental assumption of WTT} is separation of scales: the parameter of nonlinearity $0< \varepsilon= |Ak| \ll 1, \,\varepsilon \sim 10^{-2},$ must be chosen so small that only one type of resonance, i.e. three-wave resonances and if not present four-wave resonances must be taken into consideration. For small $\e$, 3WR and 4WR occur at different \emph{dynamical time scales} $t/\e$ and $t/\e^2$, respectively.

 Based on the definitions given above, and both relying on separation of scales, there are two main subjects of study in WTT:  resonance clusters formed by  resonant triads or quartets of solutions of (\ref{3R}),(\ref{3D}) connected via common modes in discrete WTT  and a stationary  solution of (\ref{3K}) in  kinetic WTT. \emph{Kinetic time scales} at which stationary spectra are forming are $t/\e^2$ (3WKE) and $t/\e^4$ (4WKE).

Laboratory experiments, \cite{erik}, and numerical simulations, \cite{AnSh} demonstrate that time scale separation breaks when $\varepsilon$ is tending to $10^{-1}$; this means that the predictions of both discrete and kinetic WTT are not applicable anymore.

CASE: In \cite{denis}, surface water waves with a wavelength of 25 cm were studied in a water flume with  dimensions 6x12x1.5 meters and compared against the predictions of kinetic WTT. The main
finding was, that in accordance with  kinetic WTT for a wide range of excitation amplitudes the energy spectrum has a power-law scaling, $k^{\nu}$, however the  exponent $\nu$ was found to be non-universal, ranging from -6.5 to about -3.5 for different levels of wave excitations. Duration of an experiment: the longest experiments took about 20 minutes (S. Lukaschuk, personal communication).

ANALYSIS: Surface water waves have a dispersion relation $\o^2=gk$, so the period of linear water waves with a wave length of $\lambda=0.25$ m
is computed as
\bea k&=&2\pi/\lambda\approx  25.12 m^{-1}  \RA \o=\sqrt{gk} \approx 15.7 sec^{-1} \nn \\ &\RA&
t=2\pi/\o \approx 0.4 sec.
\eea
Surface water waves are a 4-wave system, and for a small parameter in the magnitude of
 $\e =0.01$  the corresponding time scales can easily be computed. The dynamical time scale is of the order of  $t/\e^2 \sim $ 70 minutes  and the kinetic time scale is of the order of
$t/\e^4 \sim$ 463 days. This means that kinetic WTT can not be applied, or put it the other way, you cannot draw conclusions on this theory based on the findings of the experiment.

As an alternative, a theoretical approach which can be used for describing energy spectra in the experiments described above is the increment chain equation method recently introduced in \cite{K12a}. This method allows to compute \emph{dynamical} energy spectrum (D-spectrum)  in  wave systems with small and moderate nonlinearity, $\e =0.1 \div 0.4$ (for gravity water waves), under the effect of narrow initial excitation.

The main mechanism underlying the formation of the dynamical energy cascade is modulation instability; the corresponding time scale is $t/\e^2$. Properties of the dynamical cascade and the computation of the energy spectra for gravity water waves are described in \cite{K12b} while the general description of the time scales for kinetic and dynamical cascades in different wave systems is given in \cite{K13}.

In particular, the energy spectra in wave systems with a dispersion relation of the form $\o(\textbf{k}) \sim k^\a$ have exponential form $E_k \sim \exp({\b k})$ with a function $\b$ depending on the initial conditions (amplitude of excitation and frequency of excitation) and also on the form of the dispersion function. As $\exp({\b k}) = \sum_{n = 0}^{\infty} {(\b k)^n \over n!}$,
by special choice of initial conditions the shape of the energy spectrum can be sometimes approximated by a few or even one term having of the power-law form. For instance, for surface gravity water waves the notable Phillips spectrum $E(\o) \sim \o^{-5}$ has been obtained as a particular case of the D-spectrum for this wave system, \cite{K12a}.

\section{3- or 4-wave resonances?}
In some cases there are simple rules to determine whether it is possible at all that a particular
set of resonance conditions can be satisfied in a given physical system. For instance, it has been proven that in a 2D system with a dispersion function of the form
$\o \sim k^{\a}$  there are solutions to the three-wave resonance conditions (\ref{3R}) if and only if $\a > 1$, \cite{Ved}. A dispersion function of this special form is said to be of decay type. It was concluded, and is still widely assumed, that in any two-dimensional system with a dispersion function of decay type one has a guarantee that this is a three wave system, i.e. actually has 3 wave resonances.

CASE: As capillary waves have a dispersion function $\o \sim k^{3/2}$ which is of decay type,
 a 3-wave kinetic equation of the form (\ref{3K}) has been written in \cite{zak2} and the corresponding form of the energy spectrum was computed. Some 30 years later
 numerical simulations were conducted to establish this energy spectrum empirically, \cite{PZ99}. Instead of a continuous spectrum, it was found that the energy stayed in whatever modes were excited, although the duration of the simulations was sufficiently long. This phenomenon was called "frozen wave turbulence".

ANALYSIS: The resonance conditions (\ref{3R}) are formulated for the coordinates of wave vectors, say $k_x=2\pi m_x/L_x$, $k_y=2\pi m_y/L_y$, etc. where $L_x, L_y$  are the sizes of the interaction domain in the $x$- and $y$-directions correspondingly, and $m_x$ and $m_y$ are indexes of the Fourier harmonics, i.e. integers. After obvious normalization, without loss of generality we can regard only integer wave numbers. As it was proven in  \cite{PHD1}, in the special case of capillary water waves there are no such integer solutions to the resonance conditions (\ref{3R}).

In this case, the kinetic WTT gives no definite prediction on what would happen. Two different scenarios are possible:

 1. detuned 3-wave resonances, that is, resonances satisfying (\ref{3R}) only approximately:
  \bea \label{3R-quasi}
|\omega ({\bf k}_1) + \omega ({\bf k}_2)- \omega ({\bf k}_{3})|=\Delta \o, \
{\bf k}_1 + {\bf k}_2 ={\bf k}_{3},
\eea
with $\Delta(\o)>0$ being called resonance detuning.

2. exact 4-wave resonances.
 \bea
\omega ({\bf k}_1) + \omega ({\bf k}_2)= \omega ({\bf k}_{3}) + \omega ({\bf k}_{4}), \nn \ \
{\bf k}_1 + {\bf k}_2 ={\bf k}_{3}+{\bf k}_{4}; \label{4WR}
\eea

The numerical results presented in \cite{PZ99} reveal the emergence of rare detuned 3-wave interactions while 4-wave interactions were not taken into account in this model. The properties of the detuned interactions of capillary waves were studied in \cite{KK13}; it was shown that the minimal angle between two approximately interacting capillary waves is $78^{\circ}$, i.e. they are almost perpendicular.

This fact contradicts another assumption of the kinetic WTT that all wave vectors should be almost collinear; this assumption provides convergence of the integral in the right hand side of the kinetic  equation (\ref{3K}). Accordingly, capillary water waves must be treated as a four wave system (\ref{4WR}) when applying the kinetic WTT.

This conclusion is supported by the results of the experimental studies conducted by a few different groups of researchers where the evidence of strong four-wave coupling was found in experimental data with capillary water waves, \cite{erik,SPX10a,SPX10b,ParEx1}.

\begin{figure}
\includegraphics[width=4cm,height=6cm,angle=270]{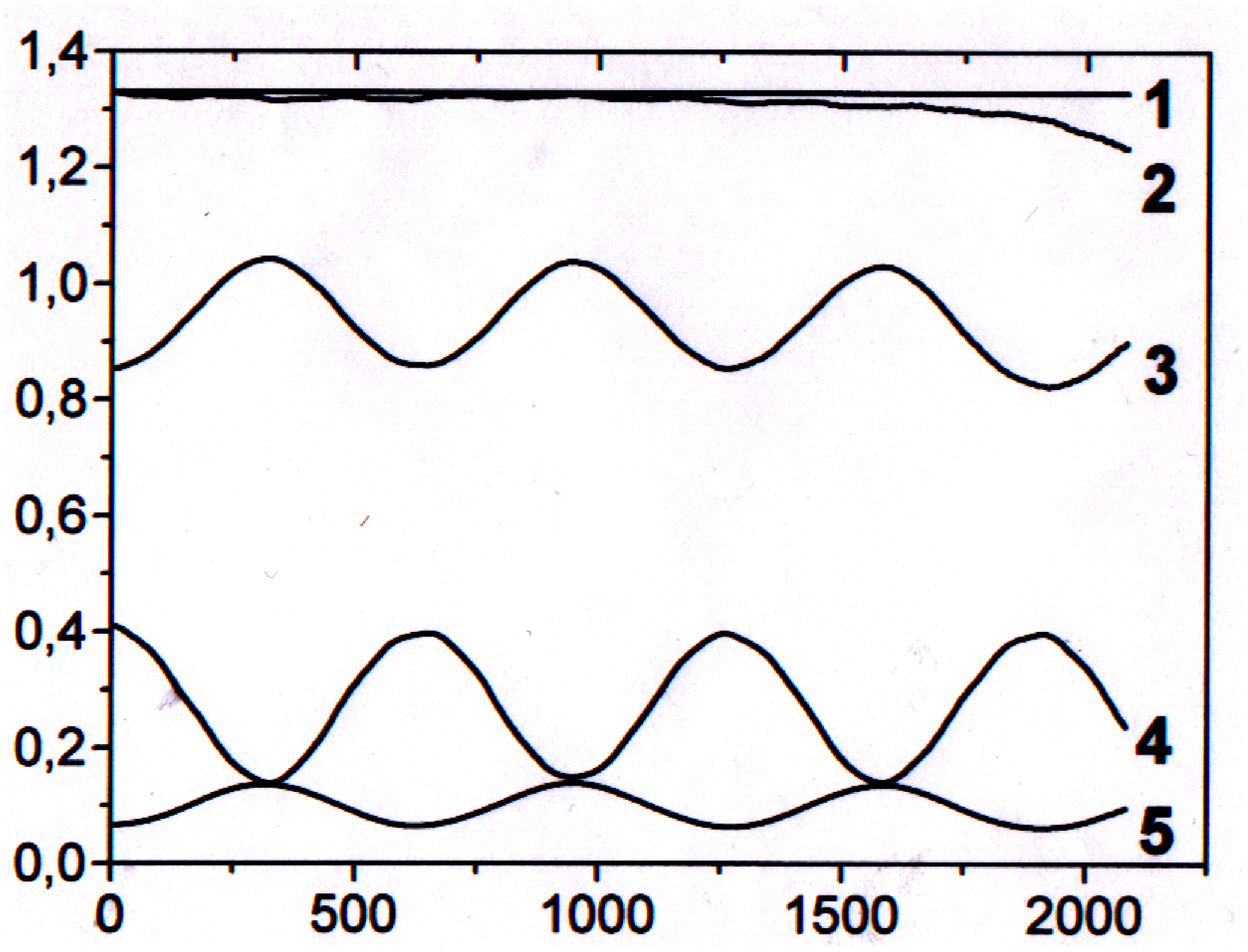}
\includegraphics[width=4cm,height=6cm,angle=270]{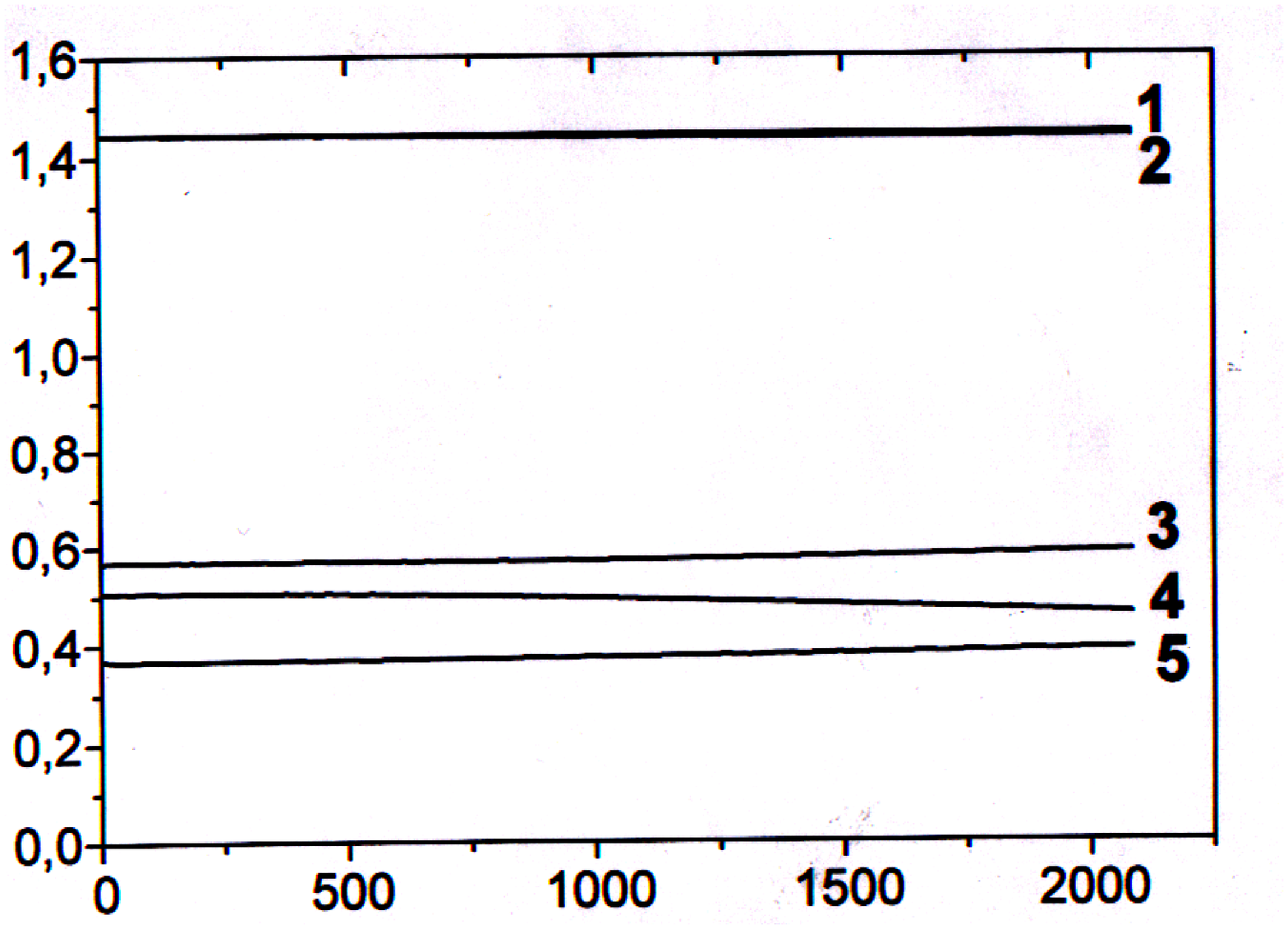}
\caption{\label{f:1} Characteristic time evolution of resonant modes (upper panel) and non-resonant modes (lower panel). Total initial energy is the same in both panels. Vertical and horizontal axes denote energy and time correspondingly, in undimensioned  units. 1- total energy of all  modes in the spectral domain studied; 2 - total energy in three modes;  2,3,4 - energy of each of three initially excited modes.}
\end{figure}

Just to complete the picture: even if there are three wave resonances, in a typical 3-wave system 40 to 60$\%$  of all modes do not take part in the exact resonances. This means that at the dynamical time scale $t/\varepsilon$ they just keep their energy and appear to be "frozen". We illustrate this with Fig.\ref{f:1} below reproduced from \cite{PRL} for reader's convenience. In this figure the results of numerical simulations with the barotropic vorticity equation on a rotating sphere are presented.

Applying this to our case, we find that the numerical results, analyzing three wave interaction and not taking into account four wave interaction, perfectly agreed with what was to be expected from discrete WTT; there is no need to introduce a new type of wave turbulence.
\section{Detuned resonances}
The discrete WTT describes the structure and dynamics of the exact integer solutions to the resonant conditions (\ref{3R}) in terms of \emph{resonance clusters}. To demonstrate how the clusters are constructed, regard spherical planetary waves with dispersion function $\o \sim m/n(n+1)$ where $m$ and $n-m$ are  longitudinal and latitudinal wave numbers correspondingly. Resonance conditions for these waves have first been found in \cite{sil} and solved in \cite{KPR,RPK93}; some explicit solutions can be found e.g. in \cite{KL07}.

\begin{figure}
\includegraphics[width=8cm,height=6cm]{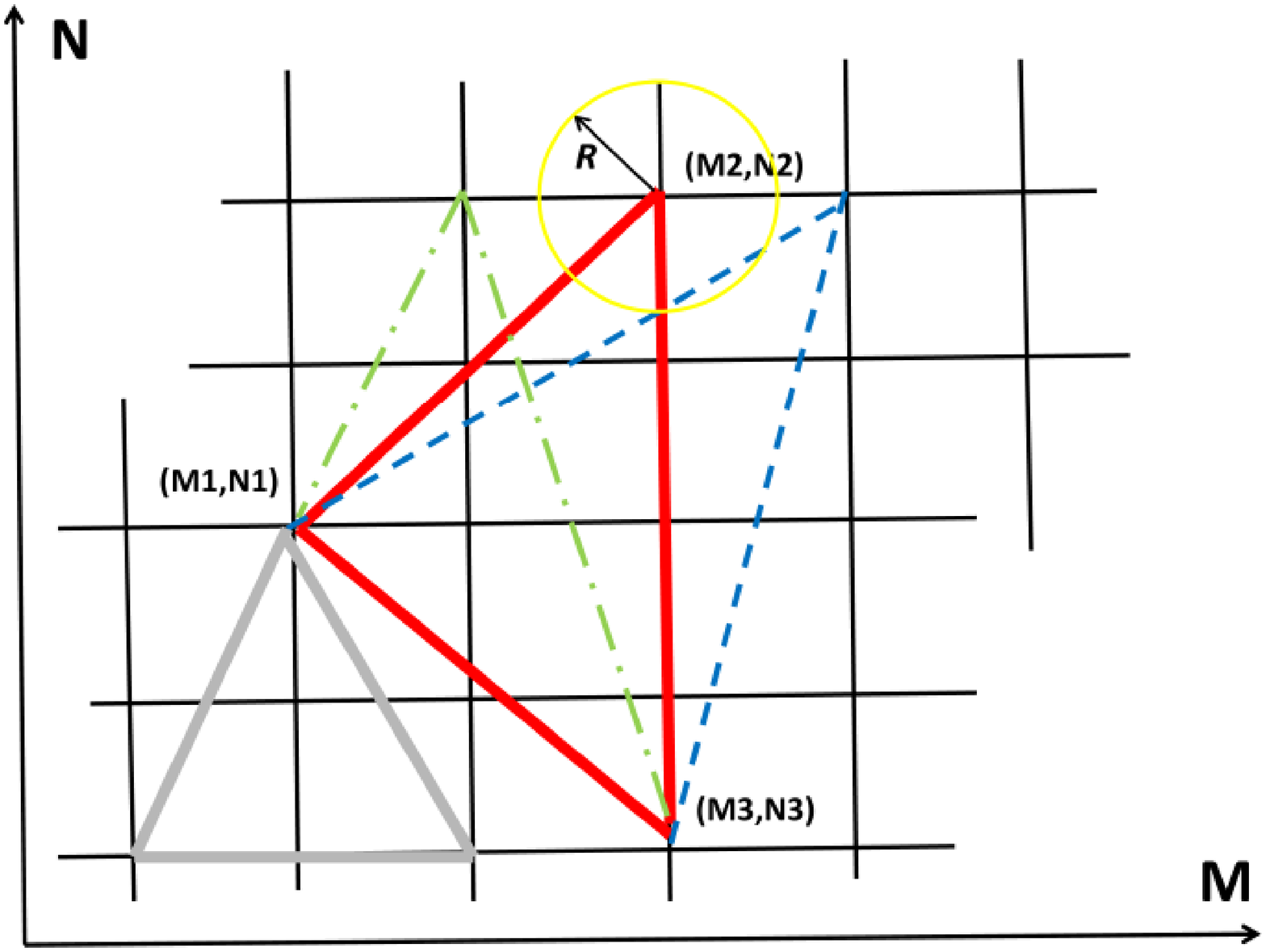}
\caption{\label{f:2} Graphical illustration of the notions of resonance cluster and  resonance detuning}
\end{figure}

In particular, it is shown that the resonant triad with three two-dimensional wave vectors ([4,12][5,14][9,13]) is isolated while the two resonant triads ([2,6][3,8][5,7]) and ([2,6][4,14][6,9]) have one joint mode [2,6]. Accordingly, they form a resonance cluster consisting of 5 resonant modes. This is illustrated schematically in  Fig.\ref{f:2} where each two-dimensional Fourier mode with wave vector $(m,n)$ is presented by a node on the integer lattice with axes $M$ and $N$. Two triangles drawn in bold red and bold grey lines with joint vertex $(m_1, n_1)$ represent a cluster of two connected resonant triads.

The dynamics of this two-triad cluster are described by two connected dynamical systems (\ref{3D}). It follows from the form of (\ref{3D}) that the dynamical system for a cluster depends on whether the connecting mode is the high-frequency mode in one or both or none of the two triads $\ a\ $ and $\ b\ $. Each case is described by  different dynamical system, namely:
\bea
\begin{cases}\label{AP}
 \dot{B}_{1a}=  Z_a B_{2a}^*B_{3a}\,, \quad  \dot{B}_{3b}=  - Z_{b} B_{3a} B_{2b}\,, \\
\dot{B}_{2a}=  Z_a B_{1a}^* B_{3a}\,,\quad    \dot{B}_{2b}=  Z_b B_{3a}^* B_{3b}\,, \\
\dot{B}_{3a}=  - Z_{a} B_{1a} B_{2a}  + Z_b B_{2b}^* B_{3b}\,,
 \end{cases}\eea \bea
 \begin{cases}\label{AA}
\dot{B}_{1a}=  Z_{a} B_{2a}^* B_{3a}\,,\    \dot{B}_{1b}=  - Z_{b}  B_{2b}^*B_{3a}\,, \\
\dot{B}_{2a}=  Z_a B_{1a}^* B_{3a}\,,\quad    \dot{B}_{2b}=  Z_b B_{1b}^* B_{3a}\,, \\
\dot{B}_{3a}= - Z_a B_{1a}B_{2a} -   Z_b B_{1b}B_{2b} \ . \\
\end{cases}\eea \bea
\begin{cases}\label{PP}
\dot{B}_{1a}=  Z_a B_{2a}^*B_{3a} +   Z_b B_{2b}^*B_{3b}\,, \\
\dot{B}_{2a}=  Z_a B_{1a}^* B_{3a}\,,\quad    \dot{B}_{2b}=  Z_b B_{1a}^* B_{3b}\,, \\
\dot{B}_{3a}=  - Z_{a} B_{1a} B_{2a} \,,\    \dot{B}_{3b}=  - Z_{b} B_{1a} B_{2b}\ . \\
\end{cases}
 \eea%
Indexes $\ a\ $ and $\ b \ $ correspond to different triads, indexes of modes "1" and "2" are interchangeable and index "3" corresponds to the high-frequency mode, e.g. $B_{3a}$ is the high-frequency mode in triad $\ a$.

This simple concept can be used for describing resonance clusters consisting of dozens or even hundreds of triads, e.g. \cite{LPPR09}, and can also be generalized for description of clusters in a 4-wave systems, \cite{CUP}.

However, this constructive approach can not be used without modification for describing clustering of detuned resonances (\ref{3R-quasi}) as it has been done in many recent publications in the field of discrete WTT. The main reason for this  is that condition (\ref{3R-quasi}) does not define a unique object. In fact,  this definition is quite open, admitting \emph{many types of solutions with completely different dynamics}.

We mention two of them, which have some prominence in literature and may be seen in analogy to resonant and non-resonant tori in KAM theory, naming them:\\
$\bullet$ \emph{quasi-resonances}: solutions for (\ref{3R-quasi}) having wave vectors with integer coordinates, and two of three modes belonging to a resonant triad.\\
From the integer nature of the wave vectors follows that detuning $\Delta \o$ can not be arbitrarily small, \cite{K07}.
The general dynamics of clusters and time scales has not been studied yet.\\
$\bullet$ \emph{non-resonant interactions}: solutions having wave vectors with real coordinates, which means that for any two arbitrarily chosen wave vectors there is a third one that (\ref{3R-quasi}) is fulfilled.\\
Detuning in this case may be arbitrarily small. The general dynamics of non-resonant triads are well understood. It has been shown that their contribution to energy exchange is negligible at time scale $t/\e$, e.g. \cite{ped,Cr85}.

The difference between these two types of detuned resonances is  illustrated schematically in the Fig.\ref{f:2}. Three wave vectors
$(m_1,n_1), (m_2,n_2), (m_3,n_3)$ satisfying exact conditions of resonance (\ref{3R}) are connected by bold red lines. Quasi-resonant triads $(m_1,n_1), (m_2-1,n_2), (m_3,n_3)$  and $(m_1,n_1), (m_2+1,n_2), (m_3,n_3)$  are shown by green dashed-dotted and blue dotted lines correspondingly. Obviously, the set of all quasi-resonances for a given resonant triad is countable (finite in the fixed spectral domain) and  is defined \emph{uniquely}.

On the other hand, detuning for non-resonant interactions can be regarded e.g. as a circle with a radius $R$ around one node of the lattice $(M,N)$ shown as a yellow circle around the node $(m_2,n_2)$. Obviously, any point on this circle gives the same resonance detuning $\Delta \o$. This means that an infinite number of waves, differing in the length of wave vector and also in the phases will produce the same $\Delta \o$, i.e. in this case \emph{no unique representation} in $k$-space exists for the set of non-resonances.

Different as they are, these two types of approximate resonances have some  properties in common: \\
1) if the dispersion  function depends on a constant, e.g. periodic Rossby waves on a $\b$-plane with $\o(\mathbf{k})=\b \, m/(m^2+n^2), \, \mathbf{k}=(m,n),$ \cite{ped}, dynamics will be determined by the interplay of three parameters - $\e$, $\Delta \o$ and $\b$; \\
2) dynamical systems will have much more complicated form including multiplicative coefficients of the form $e^{-i \Delta \o t}$ with "fast" time $t$, see \cite{RPK93}.

This means, that any conclusion about cluster dynamics, form of dynamical system, etc. made above is not valid any more. There is no general theory for describing clusters of detuned triads, and each specific case should be treated separately. In other words, kinematical and dynamical properties can be studied separately for resonance clusters but not for clusters of detuned triads.

CASE: In \cite{BH13}), an attempt was undertaken to estimate the role of quasi-resonances in the overall energy exchange of Rossby waves. The authors constructed numerous detuned triads (it cannot be concluded from the paper how they are defined), and concluded that their contribution must be considerable; the parameter $\b$  did not appear in the result.

ANALYSIS: Dealing with exact resonances, you may cancel out $\b$, but not so, as we know from above, if you are dealing with quasi-resonances. So their result cannot be correct as it does not depend on $\b$. However, this may be repaired writing  $\Delta \o/ \b$ in place of $\Delta \o$ in the result.

Much more of an issue is dynamics: for exact resonances, the dynamic system of any resonant triad may be solved and its energy transfer determined.
If we have an approximate resonance as defined above, the dynamical systems are much more complicated and no general solution is known. In some special cases we may say more:
For big $\b$, $\Delta \o/ \b$ can be made small enough and the contribution of non-resonant interactions for plane periodical Rossby waves is negligible, \cite{ped}. Moreover, as it was shown by Newell in 1960s, the time-scale at which they should appear is $C\cdot t/\e,$ with a constant $ C >1$, i.e. longer than dynamical time scale $t/\e$.

For quasi-resonances  $\Delta \o$ cannot be arbitrarily small, so  there will be no quasi-resonances in this case.  For small $\b$, without studying dynamical systems we simply do not know what will happen. This means that any conclusions on the role of detuned resonances, made without taking into account $\b$ and explicit study of dynamical systems are meaningless.
\section{Final remarks}
$\bullet$ The discrete WTT was developed for the wave systems with \emph{very small nonlinearity}, $\varepsilon \sim 10^{-2}$ and \emph{localized initial state}. Its predictions (formation of resonance clusters with known structure and dynamics) are given for dynamical time scales and can be verified experimentally, see e.g. classical laboratory studies of capillary, gravity-capillary and gravity surface water waves by the late Joe Hammack and his collaborators,
\cite{ham,HH03,HHS03}.

$\bullet$ Later experiments reported, e.g. in  \cite{erik,denis,SPX10a,SPX10b,ParEx1,NR11}, have been performed with \emph{bigger nonlinearity}, $\varepsilon \sim 10^{-1}$ and \emph{localized initial state}.  Various scenarios of the time evolution of the system were observed: $k$-spectrum consists of only discrete modes, i.e. no continuous energy spectrum is formed; $k$-spectrum has discrete and continuous part; the shape of the energy spectrum is a power law, with exponent depending on the parameters of excitation; the energy spectrum has exponential shape, etc. etc.

$\bullet$ Experimental studies of wave systems with \emph{very small nonlinearity}, $\varepsilon \sim 10^{-2}$ and \emph{distributed initial state} suitable for verifying the kinetic WTT are not known to us. It looks like this type of systems appears everywhere in nature and the kinetic WTT can be used in oceanology, magnetohydrodynamics, physics of atmosphere and many other branches of physics.

"But does the hand of wave turbulence really guide the behavior of ocean waves, capillary waves and all the examples above for which one might expect the theory to apply? Although there had been notable successes, the theory also has its limitations. One might compare its current standing, particularly with respect to experiments, to the situation regarding pattern formation in the late 1960s. By that time, there had been many theoretical advances, but the experimental confirmation of the predictions fell very much in the 'looks like' category. It took the pioneering experimental works of Ahlers, Croquette, Fauve, Gollub, Libchaber, and Swinney in the mid- to late 1970s (which overcame some extraordinary challenges of managing long-time control of external parameters) to put some of the advances on a firm footing. For wave turbulence, we are only at the beginning of the experimental stage" (p.78, \cite{NR11}). These are the last words in the excellent recent review on the kinetic WTT written in 2011 by Newell and Rump.

$\bullet$ It is difficult to make any predictions about the future technical progress in the development of experimental facilities. But if at least one of the difficulties mentioned above  - establishing a suitable distributed initial state in an experiment - would  be overcome, it would become possible to test a new constructive mathematical theory that is being developed for describing energy spectra at the dynamical time scales, \cite{Kuk06,KukShi12}, which are much shorter than the kinetic time scales of the WTT. Quite obviously, this method has great potential for future use in applications.

The main idea of this method \emph{is not} to reduce the original weakly nonlinear PDE to the wave kinetic equation as is done in WTT but to reduce it to the so-called \emph{effective PDE} which contains only resonant terms, \cite{KM-all}. The effective PDE
can be solved analytically or numerically. Preliminary results of numerical simulations with a 1-dimensional nonlinear Schr\"{o}dinger equation with random forcing clearly demonstrate the formation of an energy spectrum obeying a power law scaling predicted by the effective PDE for this case, \cite{Kuk-private}.

A similar approach has been used in  \cite{YaYo13} where an incompressible two-dimensional flow on a $\b$-plane with periodic boundary conditions is considered, and a similar effective PDE is written out.

{\textbf{Acknowledgements.}} This research has been supported by
the Austrian Science Foundation (FWF) under projects P22943-N18 and  P24671. This research was supported in part by the National Science Foundation under Grant No. NSF PHY11-25915. The author is very much obliged to the organizing committee of the program ``Wave-Flow Interaction in Geophysics, Climate, Astrophysics and Plasmas", who provided an excellent opportunity for work and discussions. The author also highly appreciates the hospitality of the Kavli Institute for Theoretical Physics (Santa Barbara, California, USA), where this work was accomplished.

\end{document}